\documentclass[]{openjournal}
\usepackage{xcolor}
\usepackage{textgreek}
\usepackage[utf8]{inputenc}
\usepackage[english]{babel}
\usepackage{amsmath}
\let\oldAA\AA
\renewcommand{\AA}{\text{\normalfont\oldAA}}

\usepackage{hyperref}
\hypersetup{
    unicode, 
    colorlinks=true,
    linkcolor=linkcolor,
    citecolor=linkcolor,
    filecolor=linkcolor,
    urlcolor=linkcolor,
}
\usepackage{color,colortbl}
\definecolor{linkcolor}{rgb}{0.0,0.3,0.5}
\usepackage{tensind}
\tensordelimiter{?}
\DeclareGraphicsExtensions{.bmp,.png,.jpg,.pdf}
\usepackage{verbatim}
\usepackage[normalem]{ulem}
\usepackage{orcidlink}
\usepackage{soul}
\begin{document}
\title{Graph-Based Light-Curve Features for Robust Transient Classification}

\author{Jesús D. Petro-Ramos$^{a}$\orcidlink{0009-0009-2407-2694}, David J. Ruiz-Morales$^{a}$\orcidlink{0009-0009-9463-4752}, and D. Sierra-Porta$^{a,*}$\orcidlink{0000-0003-3461-1347}}
\email[Jesús D. Petro-Ramos: ]{jpetro@utb.edu.co}
\email[David J. Ruiz-Morales: ]{daruiz@utb.edu.co}
\email[D. Sierra-Porta: ]{dporta@utb.edu.co}
\affiliation{Universidad Tecnológica de Bolívar. Escuela de Transformación Digital. Parque Industrial y Tecnológico Carlos Vélez Pombo Km 1 Vía Turbaco. Cartagena de Indias, 130010, Colombia}
\affiliation{$^{*}$ Corresponding author: dporta@utb.edu.co (D. Sierra-Porta)}

\begin{abstract}
We investigate graph-based representations of astronomical light curves for transient classification {on a quality-controlled, class-balanced subset of the MANTRA benchmark (minimum coverage $N_{\min}=100$ epochs; $N=1{,}705$ objects after filtering and Non--Tr. subsampling).} Each series is mapped to three visibility-graph views—horizontal (HVG), directed (DHVG), and weighted (W-HVG)—from which we extract compact, length-aware network descriptors (degree/strength moments, clustering and motifs, assortativity, path/efficiency, and spectral summaries). Using {object-level} stratified five-fold validation and tree-based learners, the best configuration (LightGBM with HVG+DHVG+W-HVG features) attains a macro–F1 of $0.622\pm0.010$ and accuracy of $0.661\pm0.010$ {on this subset}. {For context, the published MANTRA baseline reports $\mathrm{F1_{macro}}=0.528$ on the full dataset; because class priors differ after quality control, this reference is not a like-for-like comparison.} Ablations show that weighted contrasts and directed asymmetry contribute complementary gains to undirected topology. Per-class analysis highlights strong performance for CV, HPM, and Non–Tr., with residual confusions concentrated in the AGN–Blazar–SN block. These results indicate that visibility graphs offer a simple, survey-agnostic bridge between irregular photometric time series and standard classifiers, yielding competitive multiclass performance without bespoke deep architectures. We release code and feature definitions {together with the list of object IDs used in the evaluation subset} to facilitate reproducibility and future extensions.
\end{abstract}

% Write your keywords here

%\maketitle

\section{Introduction}
\label{sec:intro}
Time-domain astronomy is undergoing a fundamental transformation driven by large-scale synoptic surveys that generate massive volumes of observational data \citep{ball2009, kang2023, zuo2025}. Facilities such as the Large Synoptic Survey Telescope (LSST) will deliver millions of light curves, creating an acute need for automated methods that can distinguish diverse astrophysical sources without relying on costly spectroscopic follow-up \citep{malz2019}.

The classification problem is compounded by data-quality idiosyncrasies inherent to astronomical time series. Light curves suffer from irregular sampling, heteroscedastic uncertainties with non-uniform measurement errors, and seasonal gaps \citep{zhang2021, huijse2015}. These limitations blur morphological boundaries between classes—e.g., supernovae vs. cataclysmic variables, active galactic nuclei vs. blazars, or stellar flares vs. other transients—especially when observations are sparse or uneven \citep{zhang2021}.

Class imbalance presents an additional obstacle: scientifically interesting phenomena are typically rare relative to abundant background populations \citep{huijse2015, kang2023}. This imbalance biases learners toward majority classes and hinders the discovery of rare transients, while the scarcity of labeled training data further constrains robust model development \citep{zhang2021}.

Operational constraints raise the stakes: some events require rapid response on timescales of seconds, real-time ingestion of high-throughput alert streams, and principled handling of incomplete coverage of possible phenomena \citep{Graham2017, ball2009}. Meeting these requirements demands classifiers that are both accurate and computationally efficient, and that propagate observational uncertainty through the decision process \citep{Long2017}.

The established paradigm transforms each light curve into a vector of hand-crafted statistical descriptors—ranging from distributional moments and variability indices to autocorrelation-based measures—and then trains conventional classifiers \citep{Lo2014, Bloom2012, Nun2017}. Time-ordered metrics and frequency analysis add discriminative power for periodic sources, but frequency-domain features require substantial data and are vulnerable to aliasing under irregular cadences \citep{Bloom2012}.

Despite notable successes—e.g., decision-tree and random-forest pipelines exceeding 90\% completeness with $<10\%$ contamination for several classes \citep{Graham2017, Richards2011}—feature-based approaches face persistent challenges. They entail expensive engineering and expert knowledge to separate sub-classes in uneven, noisy, gap-ridden data \citep{Becker2020}; pipelines spend much of their budget on feature selection and fitting; classifiers struggle with underrepresented or novel classes; and degeneracies between similar light-curve shapes remain \citep{Hlo2023}.

A complementary direction preserves temporal structure by reframing each light curve as a graph whose connectivity encodes simple geometric visibility relations \citep{Blancato2022}. In the horizontal visibility graph (HVG) two observations are linked if no intervening point exceeds the lower of the pair; directed HVG (DHVG) orients edges forward in time to capture asymmetry; and weighted HVG (W-HVG) attaches edge weights that reflect amplitude contrasts or measurement uncertainties \citep{Blancato2022}. These constructions depend only on relative ordering and local visibility, making them resilient to monotone transformations and modest cadence irregularity while translating extrema, bursts, plateaus, and skewness into compact graph descriptors (degree distributions, motif profiles, clustering, assortativity, path/efficiency, spectral signatures) that integrate cleanly with standard learners \citep{Audenaert2025, Ksoll2020, Garraffo2021}. {We therefore situate our study within the MANTRA benchmark as a widely used public reference for transient classification, and we report results on a quality-controlled, class-balanced subset (defined by a minimum-epochs requirement and non-transient subsampling) to ensure stable graph construction and informative macro-averaged evaluation across eight classes \citep{Neira2020}. We evaluate a simple, reproducible pipeline that builds HVG/DHVG/W-HVG representations, extracts network features, and trains modern tree ensembles; complementary large-scale efforts in heterogeneous survey data \citep{Fei2024} and alternative representations such as dm--dt mappings with convolutional networks \citep{Mahabal2011, Mahabal2017}, as well as targeted population studies \citep{Zhang2023}, provide useful context for our approach.}

\section{Data Description}
\label{sec:data}

We base our study on the MANTRA reference dataset for astronomical transient event recognition \citep{Neira2020}. MANTRA provides a curated, labeled collection of single-band photometric light curves and a stable taxonomy intended for machine-learning research. Following MANTRA, we adopt eight classes: supernovae (SN), cataclysmic variables (CV), active galactic nuclei (AGN), high proper motion stars (HPM), blazars, stellar flares (Flare), a heterogeneous class (Other), and a non-transient class (Non-Tr.). {In this work, the evaluation is performed on a quality-controlled, class-balanced subset of MANTRA defined by a minimum-coverage requirement and a controlled Non--Tr. cap (see Table~\ref{tab:mantra_subset}). Therefore, the dataset composition differs from the full MANTRA release and published full-dataset baseline numbers should be interpreted as contextual references rather than strict like-for-like benchmarks.}

The release employed here consists of two tables. The labels table contains one row per object with the fields \texttt{ID} (integer identifier), \texttt{Classification} (one of the eight classes), and \texttt{Instances} (the number of photometric samples available for that object). The light table stores the time series in long format with the columns \texttt{ID} (object identifier), \texttt{observation\_id} (per-measurement identifier), \texttt{Mag} (magnitude as provided by MANTRA), and \texttt{Magerr} (reported photometric uncertainty). Time is represented in Modified Julian Date (MJD) and serves as the index of the light table. These components allow a direct join on \texttt{ID} to retrieve each object’s light curve together with its class label. The schema mirrors the public MANTRA documentation and preserves its per-epoch uncertainty reporting \citep{Neira2020}. {For reproducibility, we also compute the effective number of epochs per object directly from the light table after cleaning, and we use this value to enforce minimum coverage.}

As described by \citet{Neira2020}, the dataset was assembled to facilitate reproducible benchmarking: light curves and labels were harmonized into a common tabular layout, the class taxonomy was fixed to cover principal transient and variable families, and per-epoch magnitude uncertainties were retained to reflect realistic heteroscedastic noise. The design intentionally exposes challenges central to synoptic surveys, including irregular sampling, seasonal gaps, label imbalance, and overlapping morphologies, while remaining simple enough to encourage cross-method comparisons. {We preserve MANTRA’s raw photometry, uncertainties, and irregular sampling patterns; however, we apply explicit quality control and class-balancing steps (detailed below), so differences observed later may reflect both representation/modeling choices and the effective evaluation regime induced by quality control.}

From the labels table we restrict to objects whose \texttt{Classification} belongs to the eight MANTRA classes and for which a corresponding time series exists in the light table. {We then enforce a minimum-coverage criterion of $N_{\min}=100$ epochs per object (after cleaning) to ensure that visibility-graph construction and network descriptors are well posed and stable.} We discard measurements with non-finite \texttt{Mag} or \texttt{Magerr}, remove duplicated \texttt{observation\_id} within each \texttt{ID}, and strictly sort observations by MJD. When extremely long light curves are present, we optionally cap the maximum number of points per object through uniform subsampling to control computational variance; this cap is chosen conservatively so as not to alter qualitative temporal structure.

{The Non--Tr. class is overwhelmingly large in the full MANTRA release. To prevent a negative-dominated regime and to keep macro-averaged metrics informative for transient classes, we cap the effective Non--Tr. sample to a fixed size (283 objects) after applying the same $N_{\min}$ criterion. Non--Tr. light curves are distributed across multiple shards; in our data extraction step we used shards 0--2 to construct the Non--Tr. candidate pool, which are provided as randomly partitioned files in the public distribution, and we verified that the epochs-per-object distribution across these shards is consistent. Table~\ref{tab:mantra_subset} reports the resulting effective class distribution used throughout the experiments.}

{
\begin{table}[ht]
\centering
\caption{Effective MANTRA evaluation subset used in this work. Published counts correspond to the eight-class MANTRA configuration reported by \citet{Neira2020}. Our effective subset reflects (i) explicit quality control ($N_{\min}=100$ epochs after cleaning), and (ii) a controlled cap of the Non--Tr. class to avoid extreme imbalance. Positive $\Delta$ values indicate that our effective sample composition differs from the published benchmark configuration due to the combined effect of quality control and label harmonization/selection steps in our preprocessing pipeline; we therefore treat published full-dataset baselines as contextual references rather than like-for-like comparisons.}
\label{tab:mantra_subset}
\begin{tabular}{lccc}
\hline
\textbf{Class} & \textbf{MANTRA (published)} & \textbf{Ours (effective subset)} & \textbf{$\Delta$} \\
\hline
SN      & 323   & 242  & -81 \\
CV      & 215   & 386  & +171 \\
AGN     & 106   & 389  & +283 \\
HPM     & 76    & 67   & -9 \\
Blazar  & 59    & 140  & +81 \\
Flare   & 51    & 134  & +83 \\
Other   & 234   & 64   & -170 \\
Non--Tr.& 18556 & 283  & -18273 \\
\hline
Total   & 19620 & 1705 & \\
\hline
\end{tabular}
\end{table}
}

{We report class counts computed from the MANTRA files used in this study (before and after quality control). Counts reported by \citet{Neira2020} are included only as published reference values; differences may reflect dataset snapshots and/or selection criteria not identical to ours.}

All processing begins from MANTRA’s magnitude and uncertainty columns without global rescaling or detrending, beyond optional outlier mitigation applied per light curve after chronological sorting. The irregular sampling and seasonal gaps present in MANTRA are intentionally retained. Where relevant for sensitivity checks, we also consider a flux-like transform derived from magnitudes (optionally normalized by \texttt{Magerr}); these alternatives are used only for robustness and do not replace the primary magnitude-based pipeline.

Finally, the Non-Tr. label aggregates non-event light curves that act as a hard negative set, while the Other label groups sources that do not fit cleanly into the remaining categories, following the definitions in \citet{Neira2020}. Throughout this work, object identities, class assignments, and per-point photometry are taken directly from the MANTRA distribution; any derived artifacts (e.g., feature tables) are deterministic transformations of the files described above. {To facilitate exact replication of the evaluation regime, we also release the list of object IDs included in the effective subset and the scripts implementing the quality-control and selection steps.}

\section{Methods}
\label{sec:methods}

\subsection{Feature generation: visibility graphs and network descriptors}

Let $\{(t_i,x_i)\}_{i=1}^{N}$ be a real-valued time series {with $N\ge N_{\min}$ after quality control (Section~\ref{sec:data})} with strictly increasing times $t_1<\cdots<t_N$ and samples $x_i\in\mathbb{R}$. The generic (``natural'') visibility criterion between two observations \citep{Luque2009} $(t_a,x_a)$ and $(t_b,x_b)$ with $t_a<t_b$ states that they are mutually visible if every intermediate point $(t_c,x_c)$ with $t_a<t_c<t_b$ lies strictly below the straight line joining $(t_a,x_a)$ and $(t_b,x_b)$, i.e.
\begin{equation}
x_c \;<\; x_a + (x_b-x_a)\,\frac{t_c-t_a}{t_b-t_a}\qquad
\forall\, t_c\in(t_a,t_b).
\label{eq:nvg}
\end{equation}
Mapping each observation to a node and connecting every pair that satisfies
\eqref{eq:nvg} produces a visibility graph.

The \emph{horizontal visibility graph} (HVG) \citep{Luque2009,Bezsudnov2014, Gon2016}
is the monotone simplification of \eqref{eq:nvg} obtained by replacing the line-of-sight test with a horizontal barrier. Two samples at indices $i<j$ are connected if and only if all intermediate values are smaller than the minimum of the endpoints:
\begin{equation}
\{i,j\}\in E_{\mathrm{HVG}}
\quad\Longleftrightarrow\quad
x_k < \min\{x_i,x_j\},
\label{eq:hvg}
\end{equation}
for all $k=i{+}1,\dots,j{-}1$.

This construction depends only on the ordering of the $x_i$ and is therefore invariant under any strictly monotone transform $\phi(x)$; it is also agnostic to the spacing of the sampling times $\{t_i\}$. Consecutive samples are always connected ($\{i,i{+}1\}\in E_{\mathrm{HVG}}$), so the resulting graph is connected and contains the path
$1{-}2{-}\cdots{-}N$.

The \emph{directed} HVG (DHVG) \citep{Lacasa2012, andrzejewska2022assessment} encodes temporal asymmetry by orienting the same visibility relation forward in time. For $i<j$,
\begin{equation}
(i,j)\in \vec E_{\mathrm{DHVG}}
\quad\Longleftrightarrow\quad
x_k < \min\{x_i,x_j\},
\label{eq:dhvg}
\end{equation}
with the arc $i\to j$, for all $k=i{+}1,\dots,j{-}1$.

Thus every undirected HVG edge becomes a single arc pointing from the earlier to the later sample, yielding an acyclic digraph whose natural topological order is the time order. Temporal irreversibility and trend asymmetries are then reflected in statistics based on in- and out-degrees
$\bigl(k_i^{\mathrm{in}},k_i^{\mathrm{out}}\bigr)$, directed motifs, and assortativity of the directed network.

A weighted variant augments either construction without changing the visibility predicate \citep{Gao2020, Kong2021SK}. Given per-epoch uncertainties $\sigma_i$ (when available), we attach a nonnegative weight to each admissible pair
$(i,j)$:
\begin{eqnarray}
w_{ij}^{(\Delta)}=\lvert x_i-x_j\rvert,\quad
w_{ij}^{(\Delta/\sigma)}=\frac{\lvert x_i-x_j\rvert}
{\sqrt{\sigma_i^2+\sigma_j^2}},\quad
w_{ij}^{(z)}={\left|\frac{x_i-\mu}{s}-\frac{x_j-\mu}{s}\right|},
\label{eq:weights}
\end{eqnarray}
where {$(\mu,s)$} denotes a robust mean–scale estimate of $\{x_i\}$.

These choices emphasize amplitude contrast and, when uncertainties are available, incorporate measurement error, which in turn supports descriptors based on node strength and disparity in addition to the usual degree, clustering, motif, path, and spectral summaries of the graph or digraph.

\subsection{Network features} 
From each graph we extract compact descriptors designed to summarize local shape, temporal asymmetry, and weighted contrast while remaining length-aware \citep{Zou2019, Iacovacci2016, Herrera2025}:
(i) degree statistics: $\{\overline{k},\,\mathrm{std}(k),\,\mathrm{skew}(k)\}$; tail proxy via the exponential fit to $P(k)$ above $k_{\min}$;
(ii) clustering and motifs: undirected clustering coefficient mean, triangle counts, and small subgraph frequencies; for DHVG, transitivity, reciprocity, and in/out motif profiles;
(iii) assortativity (Pearson degree–degree) and directed assortativity variants;
(iv) path/efficiency: global efficiency and average shortest-path length on the giant component;
(v) spectral summaries: leading eigenvalues of adjacency and Laplacian, spectral radius, and algebraic connectivity;
(vi) weighted statistics (W-HVG): node strength $\{s_i=\sum_j w_{ij}\}$, disparity $Y_i=\sum_j (w_{ij}/s_i)^2$, weighted clustering, and strength moments.
All features are computed per light curve. Where appropriate, we normalize by $N$ or $\log N$ to reduce dependence on series length and apply winsorization only for sensitivity checks, not in the main results.

Ablation-ready feature sets. We define four disjoint inputs used in our study: HVG-only, DHVG-only, W-HVG-only, and the concatenation HVG+DHVG+W-HVG. This enables attribution analyses of where discriminative signal originates.

\subsection{Learning algorithms}
We consider tree-based learners that are strong baselines on tabular features and robust {under residual class imbalance and reweighting schemes}: Random Forest (RF) \citep{Pal2005}, Extremely Randomized Trees (ET) \citep{geurts2006extremely}, Gradient Boosting \citep{friedman2002stochastic} with LightGBM (LGBM) \citep{ke2017lightgbm}, and (optionally) XGBoost. Each model is wrapped in a preprocessing pipeline with median imputation for missing values and class reweighting via \texttt{class\_weight=balanced} (or equivalent sample weights). Hyperparameters are selected by randomized search on an inner stratified $K$-fold, using macro-F1 as the objective; typical search ranges include tree depth, leaf size, subsampling, column sampling, learning rate (for boosting), and $\ell_1/\ell_2$ regularization. We record native feature importances where exposed by the estimator and aggregate them across folds to obtain stable rankings. Probability calibration (Platt or isotonic) is considered in exploratory analyses but not used in the main decision rule, which is the multiclass argmax over predicted probabilities.

To prevent leakage, all transformations (imputation, class weighting, hyperparameter selection) are fit on training folds only. The split unit is the object identifier, keeping all epochs of a light curve within the same fold.

\subsection{Evaluation protocol}
We adopt the eight-class MANTRA taxonomy \citep{Neira2020} and evaluate models with stratified 5-fold cross-validation at the object level. The primary figure of merit is macro-averaged F1 to balance minority and majority classes; we also report overall accuracy. For diagnostic analyses we compute class-wise precision, recall, and F1, confusion matrices (normalized by true class), and, when probabilities are available, one-vs-rest precision–recall area (PR-AUC). Summary numbers are reported as mean $\pm$ standard deviation across folds; aggregate confusion matrices are obtained by averaging the per-fold matrices. All ablations (HVG-only, DHVG-only, W-HVG-only, combined) are evaluated under the same protocol {on the same effective subset} to enable like-for-like comparison {between feature sets within our study}.

{All results are reported on the quality-controlled, class-balanced MANTRA subset described in Section~\ref{sec:data}.}

\section{Results}
\label{sec:results}
{Table~\ref{tab:overall} summarizes overall performance on the quality-controlled, class-balanced MANTRA subset defined in Section~\ref{sec:data} (Table~\ref{tab:mantra_subset}). For context, we also report the published MANTRA eight-class baseline metrics from \citet{Neira2020} (and the earlier reference in \citet{DIsanto2016}), but these full-dataset numbers are not directly comparable because the effective class priors and sample composition differ after quality control and Non--Tr. capping.} Our best configuration (LGBM + HVG+DHVG+W--HVG) attains a macro--$\mathrm{F1}$ of $0.624\pm0.010$ and accuracy of $0.6612\pm0.010$ {on our subset}. {Numerically, this macro--$\mathrm{F1}$ is higher than the published MANTRA baseline ($52.79$),} while macro precision increases (from $49.12$ to $64.55$) {and macro recall decreases (from $69.60$ to $61.03$), a shift consistent with more conservative predictions (higher precision) under our evaluation regime.}

\begin{table*}[ht]
\centering
\caption{{Overall comparison (8 classes). Our results are out-of-fold (OOF) means on the quality-controlled MANTRA subset (Section~\ref{sec:data}; Table~\ref{tab:mantra_subset}). For context, we list the published MANTRA baseline metrics reported by \citet{Neira2020} (and \citet{DIsanto2016}) on the full dataset; these references are not like-for-like comparable due to different class priors and dataset composition.}}
\label{tab:overall}
\begin{tabular}{lcccc}
\hline
\textbf{Method} & \textbf{Accuracy} & \textbf{Precision}$_\mathrm{macro}$ & \textbf{Recall}$_\mathrm{macro}$ & \textbf{F1}$_\mathrm{macro}$ \\
\hline
D'Isanto \citep{DIsanto2016} & -- & 46.55\% & 66.76\% & 49.92\% \\
MANTRA (baseline) \citep{Neira2020} & -- & 49.12\% & 69.60\% & 52.79\% \\
Ours (LGBM + HVG{+}DHVG{+}W\text{-}HVG) & 66.12\% & \textbf{64.55\%} & 61.03\% & \textbf{62.43\%} \\
\hline
\end{tabular}
\end{table*}

{Table~\ref{tab:perclass} reports per-class precision/recall/$\mathrm{F1}$ for our OOF predictions on the effective subset, together with the published MANTRA per-class metrics from \citet{Neira2020} shown for contextual reference (not like-for-like).} Within our subset, we observe strong performance for \textbf{CV} ($\mathrm{F1}=79.59$), \textbf{HPM} ($\mathrm{F1}=76.06$), and \textbf{Non--Tr.} ($\mathrm{F1}=96.56$), while \textbf{AGN}, \textbf{Blazar}, and \textbf{SN} remain the most challenging group (lower recall and concentrated confusions). {Relative to the published MANTRA reference, our system emphasizes precision over recall in the AGN/Blazar/SN block; however, because the effective class priors differ after quality control, these cross-paper differences should be interpreted qualitatively rather than as strict gains/losses.}

\begin{table}[ht]
\centering
\caption{Per-class metrics (percent). {Ours are out-of-fold (OOF) results on the quality-controlled MANTRA subset (Table~\ref{tab:mantra_subset}). MANTRA values are those reported by \citet{Neira2020} on the full dataset and are provided for context (not like-for-like).} Best of the two in \textbf{bold}.}
\label{tab:perclass}
\begin{tabular}{l|cccc|cccc}
\hline
 & \multicolumn{4}{c|}{\textbf{MANTRA}} & \multicolumn{4}{c}{\textbf{Ours (LGBM + all)}} \\
\textbf{Class} & \textbf{Precision} & \textbf{Recall} & \textbf{F1} & \textbf{Cover }& \textbf{Precision} & \textbf{Recall} & \textbf{F1} & \textbf{Cover} \\
\hline
SN      & \textbf{52.91} & \textbf{56.35} & \textbf{54.57} & 323 & 39.26 & 42.04 & 40.60 & 242\\
CV      & 74.21 & 76.28 & 75.23 & 215 & \textbf{79.79} & \textbf{79.38} & \textbf{79.59} & 386\\
AGN     & \textbf{63.85} & \textbf{78.30} & \textbf{70.34} & 106 & 53.98 & 66.41 & 59.57 & 389\\
HPM     & 9.26  & \textbf{89.47} & 16.79 & 76 & \textbf{80.60} & 72.00 & \textbf{76.06} &  67\\
Blazar  & 50.82 & \textbf{52.54} & \textbf{51.67} & 59 & \textbf{51.43} & 42.35 & 46.45 & 140\\
Flare   & 11.99 & \textbf{62.75} & 20.13 & 51 & \textbf{51.49} & 47.92 & \textbf{49.64} & 134\\
Other   & 30.14 & \textbf{47.01} & 36.73 & 234 & \textbf{60.94} & 43.82 & \textbf{50.98} & 64 \\
Non--Tr.& \textbf{99.76} & 94.07 & \textbf{96.83} & 18556 & 98.94 & \textbf{94.30} & 96.56 & 283\\
\hline
\textbf{Avg/total} & 49.12 & \textbf{69.60} & 52.79 & 19620 & \textbf{64.55} & 61.03 & \textbf{62.43} & 1705\\
\hline
\end{tabular}
\end{table}

{In the per-class comparison, it is worth noting that our \emph{Cover} corresponds to the effective subset defined in Section~\ref{sec:data} and therefore differs substantially from that of \citet{Neira2020}, especially in Non--Tr. (283 vs. 18{,}556) and Other (64 vs. 234).} These differences stem from our explicit quality-control and class-balancing steps (minimum epochs per curve and a controlled Non--Tr. cap) and have two practical effects: first, by reducing the overwhelming majority of Non--Tr. the set becomes less dominated by easy negatives, which increases the informativeness of metrics such as PR--AUC and makes macro--F1 reflect behavior in minority classes; second, as the relative presence of stochastic (AGN/Blazar) and episodic (Flare) classes increases, the problem becomes more demanding for confusing pairs, as seen in the fainter diagonal of those rows. {Accordingly, the most reliable interpretation of Table~\ref{tab:perclass} is the internal error structure and class separability observed under our evaluation regime, rather than a strict cross-paper delta relative to full-dataset baselines.}

\begin{figure}[!h]
  \centering
\includegraphics[width=0.49\linewidth]{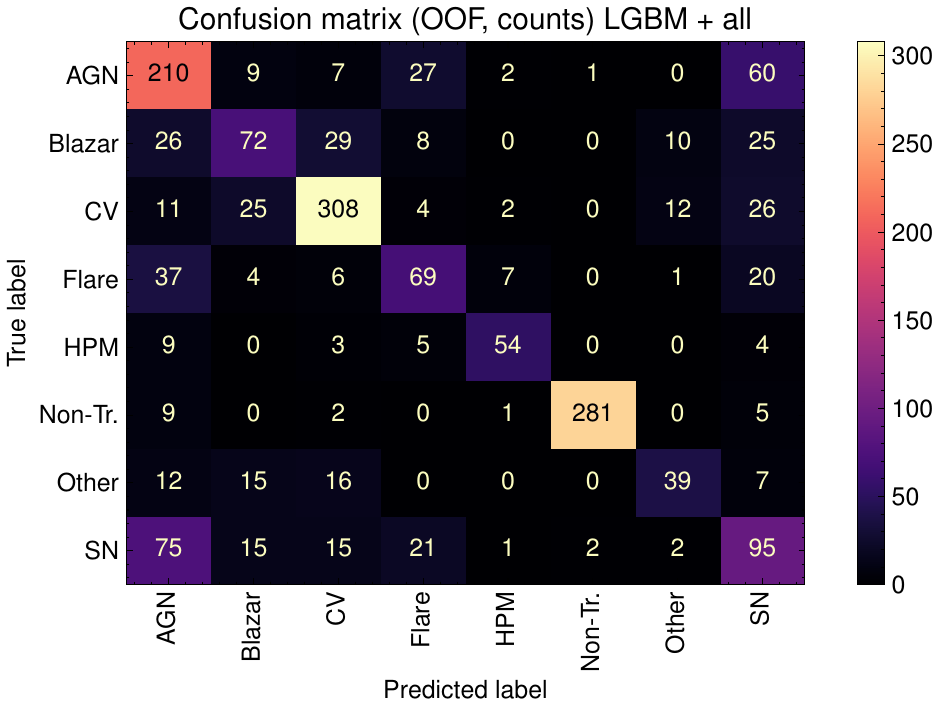}
\includegraphics[width=0.49\linewidth]{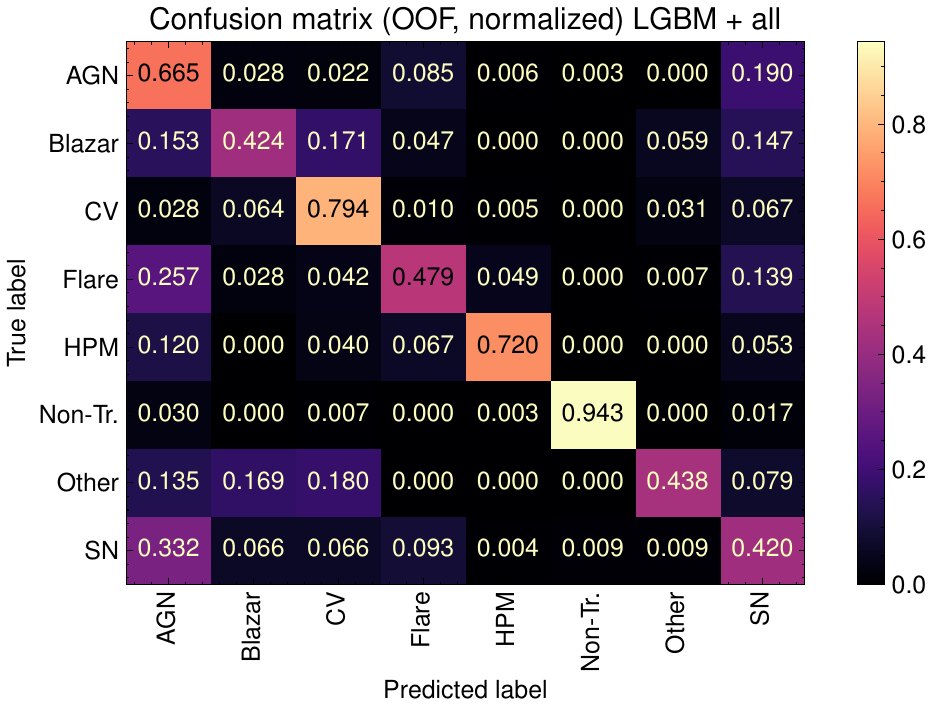}
  \caption{Out-of-fold predictions for the best configuration (LGBM + HVG+DHVG+W-HVG) {on the quality-controlled MANTRA subset} for the eight-class task.}
  \label{fig:cm}
\end{figure}

Figure~\ref{fig:cm} summarizes the error structure. In the normalized matrix (left panel) the diagonal is strongest for Non--Tr.\ and CV, followed by HPM; the hardest block spans AGN, Blazar, and SN, with confusions concentrated along their off-diagonal entries. Flare leaks primarily into AGN and SN, consistent with bursty episodes embedded in stochastic variability.

Two systematic asymmetries are worth noting. First, AGN and Blazar form a nearly symmetric confusion pair, but SN participates asymmetrically: a nontrivial fraction of SN is absorbed by AGN/Blazar, whereas the reverse is rarer. This is consistent with partial light–curve coverage: early or late SN segments without the full rise–fall morphology resemble stochastic red–noise states and are therefore pulled toward AGN/Blazar. Second, HPM shows high recall but a modest rate of false positives into CV; both classes exhibit relatively regular trends at the time scales sampled, and insufficient baseline can make monotonic drifts (HPM) appear as quasi–periodic segments (CV).

The matrix also reflects sensitivity to cadence and signal–to–noise. Rows with larger leakage often correspond to classes whose discriminative cues are concentrated in short, high–contrast windows (Flare, parts of SN). Sparse sampling or larger magnitude uncertainties dilute the weighted–contrast signal, shifting mass toward stochastic classes. Conversely, classes with distributed cues across the sequence (Non–Tr., CV) maintain strong diagonals even under irregular cadence, which aligns with the prominence of efficiency, degree–tail, and clustering features in the importance analysis.

Finally, some off–diagonal structure is consistent with boundary definitions rather than modeling capacity. Other is intentionally heterogeneous, so its errors distribute toward several neighbors; a modest gain in precision there comes at the cost of recall, which is visible as a thinner diagonal. For the AGN–Blazar–SN block, class–specific thresholds or calibrated posteriors could rebalance precision and recall without retraining: increasing the decision margin for SN reduces spurious assignments to AGN/Blazar, while per–class costs would down–weight the most common cross–confusions. Stratifying the confusion matrix by light–curve length, median Magerr, or seasonal gap metrics (not shown) leads to the same qualitative picture: improved coverage and lower uncertainty compress the off–diagonal mass in precisely those pairs where morphology is most similar under sparse sampling.

\begin{figure}[!h]
  \centering
  \includegraphics[width=0.6\linewidth]{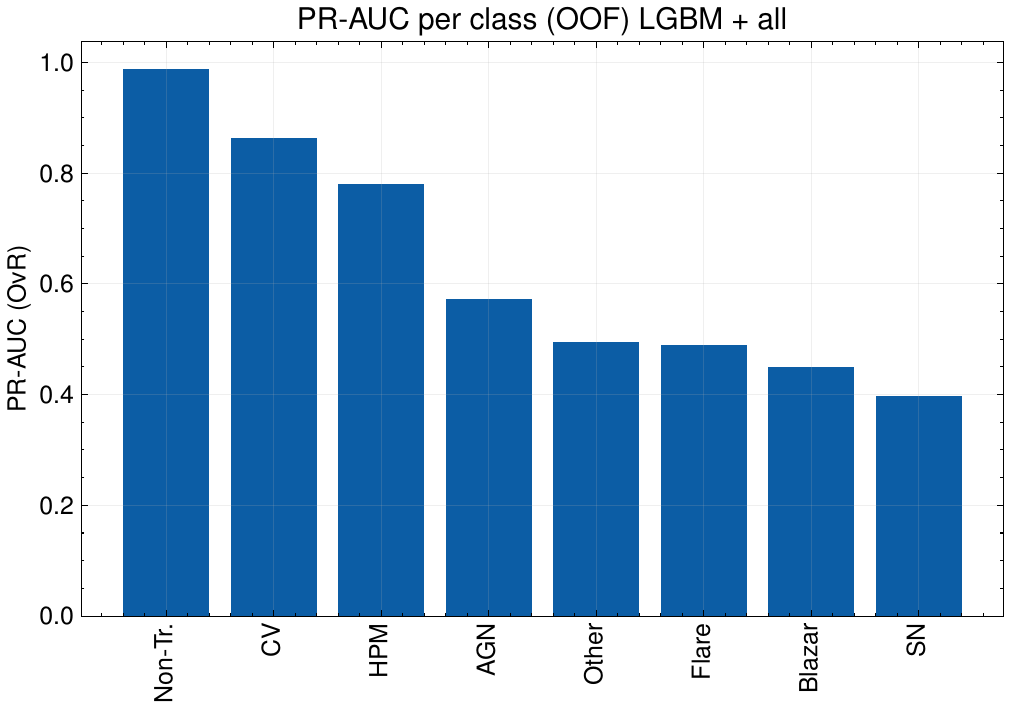}
  \caption{One-vs-rest PR–AUC per class (OOF). Higher values indicate better separability against the rest of the catalog.}
  \label{fig:prauc}
\end{figure}

Figure~\ref{fig:prauc} corroborates these patterns: Non--Tr., CV, and HPM achieve the largest PR--AUC, indicating well-separated decision regions; AGN sits mid-range; Blazar, SN, Flare, and Other are lower, reflecting overlap under sparse, irregular sampling.

Beyond ranking classes by separability, PR–AUC also exposes how base rates and calibration interact under imbalance. Because the no-skill baseline of PR–AUC equals the positive prevalence of each class, gains above that baseline are most informative in minority regimes. In our case, CV and HPM achieve large margins over their baselines, consistent with distinctive cues captured by weighted contrast (strength/disparity) and by directed asymmetry, whereas Blazar, SN, Flare, and Other show flatter precision–recall trade-offs indicative of overlapping score distributions under sparse, irregular sampling. The mid-range AGN reflects a mixture of separable episodes (e.g., long-term drifts lifting precision at moderate recall) and segments that resemble Blazar/SN. We verified that modest probability calibration shifts PR curves vertically without altering their relative ordering, suggesting that class-specific thresholding could reclaim precision for SN and Flare at small recall with limited impact on CV/HPM. Stratifying PR–AUC by light-curve length or median \texttt{Magerr} (not shown) yields the same qualitative picture: improved coverage and lower uncertainty inflate the high-precision knee primarily for CV and HPM, while stochastic classes gain more gradually across recall.

\begin{figure*}[!h]
  \centering
  \includegraphics[width=0.9\linewidth]{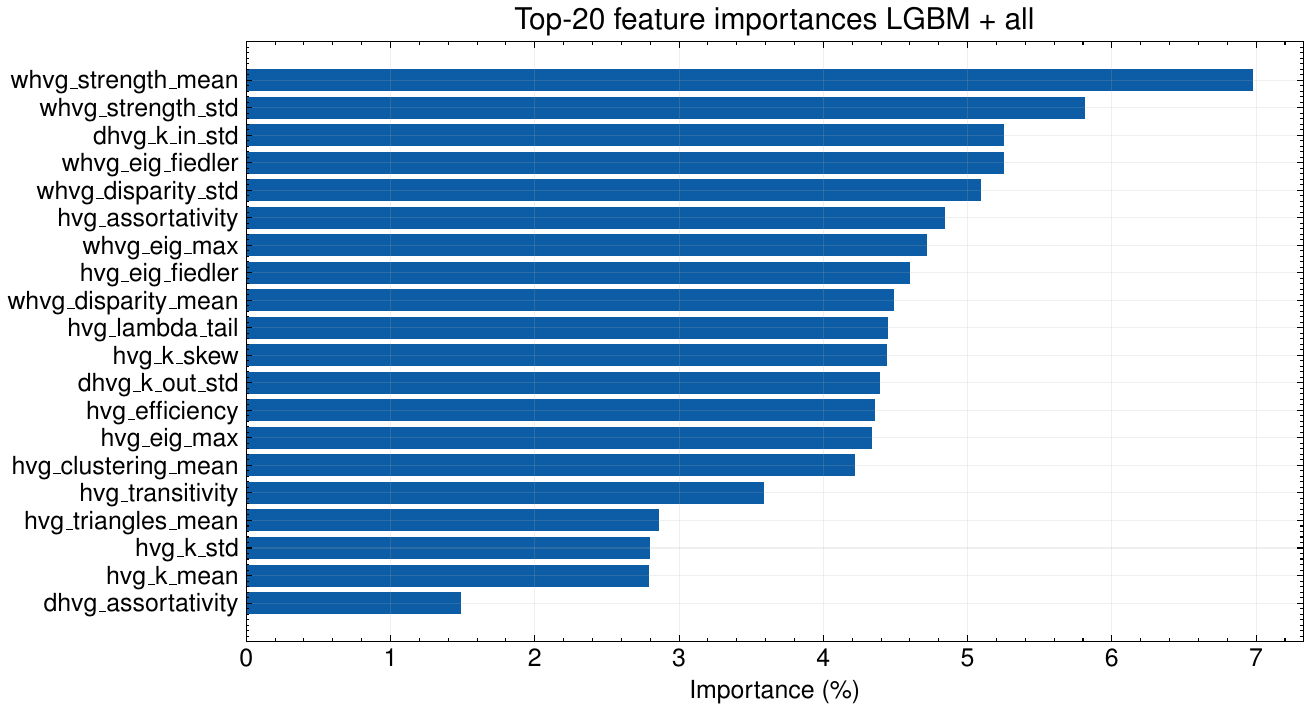}
  \caption{Model-based feature importances aggregated across folds for the best system. Bars show mean importance; whiskers show the standard deviation across folds.}
  \label{fig:featimp}
\end{figure*}

Feature attributions in Figure~\ref{fig:featimp} show a mixed signal: weighted-graph contrasts (e.g., strength and disparity from W-HVG) rank highly, directed asymmetry from DHVG contributes via in/out-degree dispersion and related motifs, and classic HVG topology (assortativity, clustering, triangles, transitivity, spectral summaries) remains informative. This suggests that amplitude-aware edges help under heteroscedastic noise, while temporal irreversibility and coarse topology provide complementary cues across classes.

{On the quality-controlled MANTRA subset, the visibility-graph representation achieves a macro--$\mathrm{F1}$ of $0.622\pm0.010$ with accuracy $0.661\pm0.010$. The published full-dataset MANTRA baseline reports $\mathrm{F1_{macro}}=0.528$ \citep{Neira2020}; because dataset composition differs after quality control, we treat this number as contextual reference rather than a strict like-for-like baseline.}

The leading block is contributed by the \emph{weighted} descriptors: \texttt{whvg\_strength\_mean} and \texttt{whvg\_strength\_std} (together $\approx 12.8\%$ of total importance), followed by \texttt{whvg\_disparity\_mean/std} and the spectral summaries \texttt{whvg\_eig\_fiedler} and \texttt{whvg\_eig\_max}. This pattern is consistent with heteroscedastic, amplitude–rich light curves in MANTRA. Node strength aggregates the visibility–contrast carried by a sample’s edges; impulsive or large–amplitude behaviour (e.g., flares, sharp rise/decline phases in SNe) yields high and variable strengths, whereas Non–Tr.\ and smoother variability produce more moderate, homogeneous values. Disparity quantifies how concentrated that contrast is on a few edges versus distributed over many, which separates burst–like morphologies (high disparity) from stochastic, red–noise–like variability in AGN/Blazar (lower disparity). At a global scale, the weighted spectral terms (spectral radius and algebraic connectivity) distinguish graphs with long “visibility bridges’’—typical of plateaus and abrupt transitions—from more locally connected structures; this helps discriminate CV/HPM (more regular, monotone segments) from AGN/Blazar.

A second, complementary block arises from \emph{undirected topology} and \emph{temporal asymmetry}. HVG measures such as \texttt{hvg\_assortativity}, \texttt{hvg\_lambda\_tail}, \texttt{hvg\_k\_skew}, \texttt{hvg\_efficiency}, \texttt{hvg\_clustering\_mean}, \texttt{hvg\_transitivity}, and \texttt{hvg\_triangles\_mean} jointly account for $\sim 35\%$ of importance. Heavy degree tails and positive skew arise when prominent extrema “see’’ many neighbours (bursty classes); higher efficiency reflects shorter paths induced by extended visibility during regular oscillations (as in CV). The DHVG features \texttt{dhvg\_k\_in\_std} and \texttt{dhvg\_k\_out\_std} ($\approx 9.6\%$ combined) capture irreversibility: sustained rise/decay phases (SNe, HPM) break the balance between incoming and outgoing links, inflating the dispersion of in/out degrees. Additional directed metrics (assortativity, efficiency, spectrum) reinforce this signal. Overall, the importance profile supports a three–part picture: weighted contrasts capture amplitude and uncertainty, directed structure captures temporal asymmetry, and HVG topology captures global shape and local closure—precisely the complementary facets that {underlie the observed macro--$\mathrm{F1}$ performance on our effective evaluation subset.}

\section{Conclusions}
\label{sec:conclusions}
We presented a reproducible pipeline that encodes light-curve geometry as visibility graphs and learns on compact network descriptors. {On the quality-controlled, class-balanced MANTRA subset (minimum coverage and Non--Tr. cap), the approach attains a macro--F1 of $0.622\pm0.010$, with the strongest gains arising when HVG, DHVG, and W-HVG features are combined. For context, the published MANTRA baseline reports $\mathrm{F1_{macro}}=0.528$ on the full dataset \citep{Neira2020}, but this reference is not like-for-like comparable due to different class priors after quality control.} Feature attributions reveal a complementary triad: weighted contrasts capture amplitude and heteroscedastic noise, directed structure captures temporal asymmetry and irreversibility, and undirected topology captures global shape and local closure. Remaining errors cluster among AGN, Blazar, and SN, suggesting future work on class-specific decision thresholds, cost-sensitive training, and refined weighting schemes tailored to stochastic versus impulsive variability. Overall, visibility-graph representations provide an effective, survey-agnostic alternative to heavy feature engineering or bespoke deep models for synoptic classification.

{
\paragraph{Limitations and outlook.}
The present study focuses on a quality-controlled, class-balanced evaluation regime to ensure stable graph construction under irregular cadences; consequently, the reported metrics characterize performance under this effective subset rather than the full MANTRA release. A natural next step is to extend the same pipeline to the complete benchmark and to additional surveys with different cadence/noise profiles, while preserving the same object-level protocol. Because the representation depends only on visibility relations and requires no domain-specific preprocessing, we anticipate that visibility-graph features can serve as a lightweight, interpretable component in larger alert-stream systems, either as a standalone tabular baseline or as complementary features alongside learned time-series embeddings.}

\section{Data and Code Availability}
\label{sec:availability}

All code used in this study is publicly available at
\url{https://github.com/JesusPetro/lightcurve-graph-features/blob/master}.
{The repository contains the end-to-end workflow to (a) construct HVG/DHVG/W-HVG representations from MANTRA light curves, (b) train and evaluate the models under the object-level cross-validation protocol used in the paper, and (c) reproduce the main figures and summary tables.} {To facilitate exact replication of our evaluation regime, we also provide the list of object IDs included in the quality-controlled subset and the scripts implementing the filtering and Non--Tr. capping described in Section~\ref{sec:data}.}

The MANTRA dataset \citep{Neira2020} is not redistributed; instructions to obtain the original labels and photometry and to run our preprocessing {from the raw MANTRA files} are provided in the README. {A minimal environment specification is included to ease installation and ensure consistent results across systems.} For archival reproducibility, we also provide the exact command lines and configuration used for the main run, {and we will report the corresponding commit hash in the camera-ready version.}

\section*{Acknowledgments}
This work was supported by the Research Directorate of Universidad Tecnológica de Bolívar (UTB), Cartagena, Colombia, which provided institutional support and encouragement during the development of this study.

\bibliographystyle{aasjournal}
\bibliography{main}{}

@ARTICLE{ball2009,
       author = {{Ball}, Nicholas M. and {Brunner}, Robert J.},
        title = "{Data Mining and Machine Learning in Astronomy}",
      journal = {International Journal of Modern Physics D},
         year = 2010,
        month = jan,
       volume = {19},
       number = {7},
        pages = {1049-1106},
          doi = {10.1142/S0218271810017160},
archivePrefix = {arXiv},
       eprint = {0906.2173},
 primaryClass = {astro-ph.IM},
       adsurl = {https://ui.adsabs.harvard.edu/abs/2010IJMPD..19.1049B}
}

@ARTICLE{kang2023,
       author = {{Kang}, Zihan and {Zhang}, Yanxia and {Zhang}, Jingyi and {Li}, Changhua and {Kong}, Minzhi and {Zhao}, Yongheng and {Wu}, Xue-Bing},
        title = "{Periodic Variable Star Classification with Deep Learning: Handling Data Imbalance in an Ensemble Augmentation Way}",
      journal = {\pasp},
         year = 2023,
        month = sep,
       volume = {135},
       number = {1051},
          eid = {094501},
        pages = {094501},
          doi = {10.1088/1538-3873/acf15e},
archivePrefix = {arXiv},
       eprint = {2309.13629},
 primaryClass = {astro-ph.IM},
       adsurl = {https://ui.adsabs.harvard.edu/abs/2023PASP..135i4501K}
}

@ARTICLE{zuo2025,
       author = {{Zuo}, Xiaoxiong and {Tao}, Yihan and {Huang}, Yang and {Kang}, Zhixuan and {Chen}, Huaxi and {Cui}, Chenzhou and {Pan}, Jiashu and {Kong}, Xiao and {Tang}, Xiaoyu and {Han}, Henggeng and {Mu}, Haiyang and {Xu}, Yunfei and {Fan}, Dongwei and {Xue}, Guirong and {Luo}, Ali and {Liu}, Jifeng},
        title = "{FALCO: a Foundation model of Astronomical Light Curves for time dOmain astronomy}",
      journal = {arXiv e-prints},
         year = 2025,
        month = apr,
          eid = {arXiv:2504.20290},
        pages = {arXiv:2504.20290},
          doi = {10.48550/arXiv.2504.20290},
archivePrefix = {arXiv},
       eprint = {2504.20290},
 primaryClass = {astro-ph.IM},
       adsurl = {https://ui.adsabs.harvard.edu/abs/2025arXiv250420290Z}
}

@ARTICLE{malz2019,
       author = {{Malz}, A.~I. and {Hlo{\v{z}}ek}, R. and {Allam}, Jr., T. and {Bahmanyar}, A. and {Biswas}, R. and {Dai}, M. and {Galbany}, L. and {Ishida}, E.~E.~O. and {Jha}, S.~W. and {Jones}, D.~O. and {Kessler}, R. and {Lochner}, M. and {Mahabal}, A.~A. and {Mandel}, K.~S. and {Mart{\'\i}nez-Galarza}, J.~R. and {McEwen}, J.~D. and {Muthukrishna}, D. and {Narayan}, G. and {Peiris}, H. and {Peters}, C.~M. and {Ponder}, K. and {Setzer}, C.~N. and {(the LSST Dark Energy Science Collaboration} and {LSST Transients}, the and {Variable Stars Science Collaboration}},
        title = "{The Photometric LSST Astronomical Time-series Classification Challenge PLAsTiCC: Selection of a Performance Metric for Classification Probabilities Balancing Diverse Science Goals}",
      journal = {\aj},
         year = 2019,
        month = nov,
       volume = {158},
       number = {5},
          eid = {171},
        pages = {171},
          doi = {10.3847/1538-3881/ab3a2f},
archivePrefix = {arXiv},
       eprint = {1809.11145},
 primaryClass = {astro-ph.IM},
       adsurl = {https://ui.adsabs.harvard.edu/abs/2019AJ....158..171M}
}

@ARTICLE{zhang2021,
       author = {{Zhang}, Keming and {Bloom}, Joshua S.},
        title = "{Classification of periodic variable stars with novel cyclic-permutation invariant neural networks}",
      journal = {\mnras},
         year = 2021,
        month = jul,
       volume = {505},
       number = {1},
        pages = {515-522},
          doi = {10.1093/mnras/stab1248},
archivePrefix = {arXiv},
       eprint = {2011.01243},
 primaryClass = {astro-ph.IM},
       adsurl = {https://ui.adsabs.harvard.edu/abs/2021MNRAS.505..515Z}
}

@ARTICLE{huijse2015,
       author = {{Huijse}, Pablo and {Estevez}, Pablo A. and {Protopapas}, Pavlos and {Principe}, Jose C. and {Zegers}, Pablo},
        title = "{Computational Intelligence Challenges and Applications on Large-Scale Astronomical Time Series Databases}",
      journal = {arXiv e-prints},
         year = 2015,
        month = sep,
          eid = {arXiv:1509.07823},
        pages = {arXiv:1509.07823},
          doi = {10.48550/arXiv.1509.07823},
archivePrefix = {arXiv},
       eprint = {1509.07823},
 primaryClass = {astro-ph.IM},
       adsurl = {https://ui.adsabs.harvard.edu/abs/2015arXiv150907823H}
}

@INPROCEEDINGS{Graham2017,
       author = {{Graham}, Matthew and {Drake}, Andrew and {Djorgovski}, S.~G. and {Mahabal}, Ashish and {Donalek}, Ciro},
        title = "{Challenges in the automated classification of variable stars in large databases}",
    booktitle = {European Physical Journal Web of Conferences},
         year = 2017,
       series = {European Physical Journal Web of Conferences},
       volume = {152},
        month = sep,
    publisher = {EDP},
          eid = {03001},
        pages = {03001},
          doi = {10.1051/epjconf/201715203001},
       adsurl = {https://ui.adsabs.harvard.edu/abs/2017EPJWC.15203001G}
}

@ARTICLE{Long2017,
       author = {{Long}, James P. and {de Souza}, Rafael S.},
        title = "{Statistical methods in astronomy}",
      journal = {arXiv e-prints},
         year = 2017,
        month = jul,
          eid = {arXiv:1707.05834},
        pages = {arXiv:1707.05834},
          doi = {10.48550/arXiv.1707.05834},
archivePrefix = {arXiv},
       eprint = {1707.05834},
 primaryClass = {physics.ed-ph},
       adsurl = {https://ui.adsabs.harvard.edu/abs/2017arXiv170705834L}
}

@ARTICLE{Lo2014,
       author = {{Lo}, Kitty K. and {Farrell}, Sean and {Murphy}, Tara and {Gaensler}, B.~M.},
        title = "{Automatic Classification of Time-variable X-Ray Sources}",
      journal = {\apj},
         year = 2014,
        month = may,
       volume = {786},
       number = {1},
          eid = {20},
        pages = {20},
          doi = {10.1088/0004-637X/786/1/20},
archivePrefix = {arXiv},
       eprint = {1403.0188},
 primaryClass = {astro-ph.IM},
       adsurl = {https://ui.adsabs.harvard.edu/abs/2014ApJ...786...20L}
}

@INCOLLECTION{Bloom2012,
       author = {{Bloom}, Joshua S. and {Richards}, Joseph W.},
        title = "{Data Mining and Machine Learning in Time-Domain Discovery and Classification}",
    booktitle = {Advances in Machine Learning and Data Mining for Astronomy},
         year = 2012,
       editor = {{Way}, Michael J. and {Scargle}, Jeffrey D. and {Ali}, Kamal M. and {Srivastava}, Ashok N.},
        pages = {89-112},
          doi = {10.48550/arXiv.1104.3142},
       adsurl = {https://ui.adsabs.harvard.edu/abs/2012amld.book...89B}
}

@software{Nun2017,
       author = {{Nun}, Isadora and {Protopapas}, Pavlos and {Sim}, Brandon and {Zhu}, Ming and {Dave}, Rahul and {Castro}, Nicolas and {Pichara}, Karim},
        title = "{FATS: Feature Analysis for Time Series}",
 howpublished = {Astrophysics Source Code Library, record ascl:1711.017},
         year = 2017,
        month = nov,
          eid = {ascl:1711.017},
archivePrefix = {ascl},
       eprint = {1711.017},
       adsurl = {https://ui.adsabs.harvard.edu/abs/2017ascl.soft11017N}
}

@ARTICLE{Richards2011,
       author = {{Richards}, Joseph W. and {Starr}, Dan L. and {Butler}, Nathaniel R. and {Bloom}, Joshua S. and {Brewer}, John M. and {Crellin-Quick}, Arien and {Higgins}, Justin and {Kennedy}, Rachel and {Rischard}, Maxime},
        title = "{On Machine-learned Classification of Variable Stars with Sparse and Noisy Time-series Data}",
      journal = {\apj},
         year = 2011,
        month = may,
       volume = {733},
       number = {1},
          eid = {10},
        pages = {10},
          doi = {10.1088/0004-637X/733/1/10},
archivePrefix = {arXiv},
       eprint = {1101.1959},
 primaryClass = {astro-ph.IM},
       adsurl = {https://ui.adsabs.harvard.edu/abs/2011ApJ...733...10R}
}

@ARTICLE{Becker2020,
       author = {{Becker}, I. and {Pichara}, K. and {Catelan}, M. and {Protopapas}, P. and {Aguirre}, C. and {Nikzat}, F.},
        title = "{Scalable end-to-end recurrent neural network for variable star classification}",
      journal = {\mnras},
         year = 2020,
        month = apr,
       volume = {493},
       number = {2},
        pages = {2981-2995},
          doi = {10.1093/mnras/staa350},
archivePrefix = {arXiv},
       eprint = {2002.00994},
 primaryClass = {astro-ph.IM},
       adsurl = {https://ui.adsabs.harvard.edu/abs/2020MNRAS.493.2981B}
}

@ARTICLE{Hlo2023,
       author = {{Hlo{\v{z}}ek}, R. and {Malz}, A.~I. and {Ponder}, K.~A. and {Dai}, M. and {Narayan}, G. and {Ishida}, E.~E.~O. and {Allam}, Jr., T. and {Bahmanyar}, A. and {Bi}, X. and {Biswas}, R. and {Boone}, K. and {Chen}, S. and {Du}, N. and {Erdem}, A. and {Galbany}, L. and {Garreta}, A. and {Jha}, S.~W. and {Jones}, D.~O. and {Kessler}, R. and {Lin}, M. and {Liu}, J. and {Lochner}, M. and {Mahabal}, A.~A. and {Mandel}, K.~S. and {Margolis}, P. and {Mart{\'\i}nez-Galarza}, J.~R. and {McEwen}, J.~D. and {Muthukrishna}, D. and {Nakatsuka}, Y. and {Noumi}, T. and {Oya}, T. and {Peiris}, H.~V. and {Peters}, C.~M. and {Puget}, J.~F. and {Setzer}, C.~N. and {Siddhartha} and {Stefanov}, S. and {Xie}, T. and {Yan}, L. and {Yeh}, K. -H. and {Zuo}, W.},
        title = "{Results of the Photometric LSST Astronomical Time-series Classification Challenge (PLAsTiCC)}",
      journal = {\apjs},
         year = 2023,
        month = aug,
       volume = {267},
       number = {2},
          eid = {25},
        pages = {25},
          doi = {10.3847/1538-4365/accd6a},
archivePrefix = {arXiv},
       eprint = {2012.12392},
 primaryClass = {astro-ph.IM},
       adsurl = {https://ui.adsabs.harvard.edu/abs/2023ApJS..267...25H}
}

@ARTICLE{Blancato2022,
       author = {{Blancato}, Kirsten and {Ness}, Melissa K. and {Huber}, Daniel and {Lu}, Yuxi(Lucy) and {Angus}, Ruth},
        title = "{Data-driven Derivation of Stellar Properties from Photometric Time Series Data Using Convolutional Neural Networks}",
      journal = {\apj},
         year = 2022,
        month = jul,
       volume = {933},
       number = {2},
          eid = {241},
        pages = {241},
          doi = {10.3847/1538-4357/ac7563},
       adsurl = {https://ui.adsabs.harvard.edu/abs/2022ApJ...933..241B}
}

@ARTICLE{Garraffo2021,
       author = {{Garraffo}, Cecilia and {Protopapas}, Pavlos and {Drake}, Jeremy J. and {Becker}, Ignacio and {Cargile}, Phillip},
        title = "{StelNet: Hierarchical Neural Network for Automatic Inference in Stellar Characterization}",
      journal = {\aj},
         year = 2021,
        month = oct,
       volume = {162},
       number = {4},
          eid = {157},
        pages = {157},
          doi = {10.3847/1538-3881/ac0ef0},
archivePrefix = {arXiv},
       eprint = {2106.07655},
 primaryClass = {astro-ph.SR},
       adsurl = {https://ui.adsabs.harvard.edu/abs/2021AJ....162..157G}
}

@ARTICLE{Ksoll2020,
       author = {{Ksoll}, Victor F. and {Ardizzone}, Lynton and {Klessen}, Ralf and {Koethe}, Ullrich and {Sabbi}, Elena and {Robberto}, Massimo and {Gouliermis}, Dimitrios and {Rother}, Carsten and {Zeidler}, Peter and {Gennaro}, Mario},
        title = "{Stellar parameter determination from photometry using invertible neural networks}",
      journal = {\mnras},
         year = 2020,
        month = dec,
       volume = {499},
       number = {4},
        pages = {5447-5485},
          doi = {10.1093/mnras/staa2931},
archivePrefix = {arXiv},
       eprint = {2007.08391},
 primaryClass = {astro-ph.SR},
       adsurl = {https://ui.adsabs.harvard.edu/abs/2020MNRAS.499.5447K}
}

@ARTICLE{Audenaert2025,
       author = {{Audenaert}, Jeroen},
        title = "{From stellar light to astrophysical insight: automating variable star research with machine learning}",
      journal = {\apss},
         year = 2025,
        month = jul,
       volume = {370},
       number = {7},
          eid = {72},
        pages = {72},
          doi = {10.1007/s10509-025-04460-5},
archivePrefix = {arXiv},
       eprint = {2507.03093},
 primaryClass = {astro-ph.IM},
       adsurl = {https://ui.adsabs.harvard.edu/abs/2025Ap&SS.370...72A}
}

@ARTICLE{Neira2020,
       author = {{Neira}, Mauricio and {G{\'o}mez}, Catalina and {Su{\'a}rez-P{\'e}rez}, John F. and {G{\'o}mez}, Diego A. and {Reyes}, Juan Pablo and {Hoyos}, Marcela Hern{\'a}ndez and {Arbel{\'a}ez}, Pablo and {Forero-Romero}, Jaime E.},
        title = "{MANTRA: A Machine-learning Reference Light-curve Data Set for Astronomical Transient Event Recognition}",
      journal = {\apjs},
         year = 2020,
        month = sep,
       volume = {250},
       number = {1},
          eid = {11},
        pages = {11},
          doi = {10.3847/1538-4365/aba267},
archivePrefix = {arXiv},
       eprint = {2006.13163},
 primaryClass = {astro-ph.IM},
       adsurl = {https://ui.adsabs.harvard.edu/abs/2020ApJS..250...11N}
}

@ARTICLE{Fei2024,
       author = {{Fei}, Ya and {Yu}, Ce and {Li}, Kun and {Chen}, Xiaodian and {Zhang}, Yajie and {Cui}, Chenzhou and {Xiao}, Jian and {Xu}, Yunfei and {Tao}, Yihan},
        title = "{LEAVES: An Expandable Light-curve Data Set for Automatic Classification of Variable Stars}",
      journal = {\apjs},
         year = 2024,
        month = nov,
       volume = {275},
       number = {1},
          eid = {10},
        pages = {10},
          doi = {10.3847/1538-4365/ad785b},
       adsurl = {https://ui.adsabs.harvard.edu/abs/2024ApJS..275...10F}
}

@ARTICLE{Mahabal2017,
       author = {{Mahabal}, Ashish and {Sheth}, Kshiteej and {Gieseke}, Fabian and {Pai}, Akshay and {Djorgovski}, S. George and {Drake}, Andrew and {Graham}, Matthew and {the CSS/CRTS/PTF Collaboration}},
        title = "{Deep-Learnt Classification of Light Curves}",
      journal = {arXiv e-prints},
         year = 2017,
        month = sep,
          eid = {arXiv:1709.06257},
        pages = {arXiv:1709.06257},
          doi = {10.48550/arXiv.1709.06257},
archivePrefix = {arXiv},
       eprint = {1709.06257},
 primaryClass = {astro-ph.IM},
       adsurl = {https://ui.adsabs.harvard.edu/abs/2017arXiv170906257M}
}

@ARTICLE{Mahabal2011,
       author = {{Mahabal}, A.~A. and {Djorgovski}, S.~G. and {Drake}, A.~J. and {Donalek}, C. and {Graham}, M.~J. and {Williams}, R.~D. and {Chen}, Y. and {Moghaddam}, B. and {Turmon}, M. and {Beshore}, E. and {Larson}, S.},
        title = "{Discovery, classification, and scientific exploration of transient events from the Catalina Real-time Transient Survey}",
      journal = {Bulletin of the Astronomical Society of India},
         year = 2011,
        month = sep,
       volume = {39},
       number = {3},
        pages = {387-408},
          doi = {10.48550/arXiv.1111.0313},
archivePrefix = {arXiv},
       eprint = {1111.0313},
 primaryClass = {astro-ph.IM},
       adsurl = {https://ui.adsabs.harvard.edu/abs/2011BASI...39..387M}
}

@ARTICLE{Zhang2023,
       author = {{Zhang}, Jingyi and {Zhang}, Yanxia and {Kang}, Zihan and {Li}, Changhua and {Zhao}, Yongheng},
        title = "{A Catalog of Young Stellar Objects from the LAMOST and ZTF Surveys}",
      journal = {\apjs},
         year = 2023,
        month = jul,
       volume = {267},
       number = {1},
          eid = {7},
        pages = {7},
          doi = {10.3847/1538-4365/acd84b},
       adsurl = {https://ui.adsabs.harvard.edu/abs/2023ApJS..267....7Z}
}

@ARTICLE{Luque2009,
       author = {{Luque}, B. and {Lacasa}, L. and {Ballesteros}, F. and {Luque}, J.},
        title = "{Horizontal visibility graphs: Exact results for random time series}",
      journal = {\pre},
         year = 2009,
        month = oct,
       volume = {80},
       number = {4},
          eid = {046103},
        pages = {046103},
          doi = {10.1103/PhysRevE.80.046103},
archivePrefix = {arXiv},
       eprint = {1002.4526},
 primaryClass = {physics.data-an},
       adsurl = {https://ui.adsabs.harvard.edu/abs/2009PhRvE..80d6103L}
}

@ARTICLE{Bezsudnov2014,
       author = {{Bezsudnov}, I.~V. and {Snarskii}, A.~A.},
        title = "{From the time series to the complex networks: The parametric natural visibility graph}",
      journal = {Physica A Statistical Mechanics and its Applications},
         year = 2014,
        month = nov,
       volume = {414},
        pages = {53-60},
          doi = {10.1016/j.physa.2014.07.002},
archivePrefix = {arXiv},
       eprint = {1208.6365},
 primaryClass = {physics.data-an},
       adsurl = {https://ui.adsabs.harvard.edu/abs/2014PhyA..414...53B}
}

@ARTICLE{Gon2016,
       author = {{Gon{\c{c}}alves}, Bruna Amin and {Carpi}, Laura and {Rosso}, Osvaldo A. and {Ravetti}, Mart{\'\i}n G.},
        title = "{Time series characterization via horizontal visibility graph and Information Theory}",
      journal = {Physica A Statistical Mechanics and its Applications},
         year = 2016,
        month = dec,
       volume = {464},
        pages = {93-102},
          doi = {10.1016/j.physa.2016.07.063},
       adsurl = {https://ui.adsabs.harvard.edu/abs/2016PhyA..464...93G}
}

@ARTICLE{Lacasa2012,
       author = {{Lacasa}, L. and {Nu{\~n}ez}, A. and {Rold{\'a}n}, {\'E}. and {Parrondo}, J.~M.~R. and {Luque}, B.},
        title = "{Time series irreversibility: a visibility graph approach}",
      journal = {European Physical Journal B},
         year = 2012,
        month = jun,
       volume = {85},
       number = {6},
          eid = {217},
        pages = {217},
          doi = {10.1140/epjb/e2012-20809-8},
archivePrefix = {arXiv},
       eprint = {1108.1691},
 primaryClass = {physics.data-an},
       adsurl = {https://ui.adsabs.harvard.edu/abs/2012EPJB...85..217L}
}

@article{andrzejewska2022assessment,
  title={Assessment of time irreversibility in a time series using visibility graphs},
  author={Andrzejewska, Ma{\l}gorzata and {\.Z}ebrowski, Jan J and Rams, Karolina and Ozimek, Mateusz and Baranowski, Rafa{\l}},
  journal={Frontiers in Network Physiology},
  volume={2},
  pages={877474},
  year={2022},
  publisher={Frontiers Media SA},
doi={https://doi.org/10.3389/fnetp.2022.877474}
}

@ARTICLE{Gao2020,
       author = {{Gao}, Yiyuan and {Yu}, Dejie and {Wang}, Haojiang},
        title = "{Fault diagnosis of rolling bearings using weighted horizontal visibility graph and graph Fourier transform}",
      journal = {Measurement},
         year = 2020,
        month = jan,
       volume = {149},
          eid = {107036},
        pages = {107036},
          doi = {10.1016/j.measurement.2019.107036},
       adsurl = {https://ui.adsabs.harvard.edu/abs/2020Meas..14907036G}
}

@ARTICLE{Kong2021SK,
       author = {{Kong}, Tianjiao and {Shao}, Jie and {Hu}, Jiuyuan and {Yang}, Xin and {Yang}, Shiyiling and {Malekian}, Reza},
        title = "{EEG-Based Emotion Recognition Using an Improved Weighted Horizontal Visibility Graph}",
      journal = {Sensors},
         year = 2021,
        month = mar,
       volume = {21},
       number = {5},
          eid = {1870},
        pages = {1870},
          doi = {10.3390/s21051870},
       adsurl = {https://ui.adsabs.harvard.edu/abs/2021Senso..21.1870K}
}

@ARTICLE{Pal2005,
       author = {{Pal}, M.},
        title = "{Random forest classifier for remote sensing classification}",
      journal = {International Journal of Remote Sensing},
         year = 2005,
        month = jan,
       volume = {26},
       number = {1},
        pages = {217-222},
          doi = {10.1080/01431160412331269698},
       adsurl = {https://ui.adsabs.harvard.edu/abs/2005IJRS...26..217P}
}

@article{geurts2006extremely,
  title={Extremely randomized trees},
  author={Geurts, Pierre and Ernst, Damien and Wehenkel, Louis},
  journal={Machine learning},
  volume={63},
  number={1},
  pages={3--42},
  year={2006},
  publisher={Springer},
doi={https://doi.org/10.1007/s10994-006-6226-1}
}

@article{friedman2002stochastic,
  title={Stochastic gradient boosting},
  author={Friedman, Jerome H},
  journal={Computational statistics \& data analysis},
  volume={38},
  number={4},
  pages={367--378},
  year={2002},
  publisher={Elsevier},
doi={https://doi.org/10.1016/S0167-9473(01)00065-2}
}

@article{ke2017lightgbm,
  title={Lightgbm: A highly efficient gradient boosting decision tree},
  author={Ke, Guolin and Meng, Qi and Finley, Thomas and Wang, Taifeng and Chen, Wei and Ma, Weidong and Ye, Qiwei and Liu, Tie-Yan},
  journal={Advances in neural information processing systems},
  volume={30},
  year={2017},
url={https://proceedings.neurips.cc/paper/2017/hash/6449f44a102fde848669bdd9eb6b76fa-Abstract.html}
}

@ARTICLE{Zou2019,
       author = {{Zou}, Yong and {Donner}, Reik V. and {Marwan}, Norbert and {Donges}, Jonathan F. and {Kurths}, J{\"u}rgen},
        title = "{Complex network approaches to nonlinear time series analysis}",
      journal = {\physrep},
         year = 2019,
        month = jan,
       volume = {787},
        pages = {1-97},
          doi = {10.1016/j.physrep.2018.10.005},
archivePrefix = {arXiv},
       eprint = {2501.18737},
 primaryClass = {physics.data-an},
       adsurl = {https://ui.adsabs.harvard.edu/abs/2019PhR...787....1Z}
}

@ARTICLE{Iacovacci2016,
       author = {{Iacovacci}, Jacopo and {Lacasa}, Lucas},
        title = "{Sequential motif profile of natural visibility graphs}",
      journal = {\pre},
         year = 2016,
        month = nov,
       volume = {94},
       number = {5},
          eid = {052309},
        pages = {052309},
          doi = {10.1103/PhysRevE.94.052309},
archivePrefix = {arXiv},
       eprint = {1605.02645},
 primaryClass = {physics.data-an},
       adsurl = {https://ui.adsabs.harvard.edu/abs/2016PhRvE..94e2309I}
}

@ARTICLE{Herrera2025,
       author = {{Herrera-Acevedo}, D.~D. and {Sierra-Porta}, D.},
        title = "{Network structure and urban mobility sustainability: A topological analysis of cities from the urban mobility readiness index}",
      journal = {Sustainable Cities and Society},
         year = 2025,
        month = feb,
       volume = {119},
          eid = {106076},
        pages = {106076},
          doi = {10.1016/j.scs.2024.106076},
       adsurl = {https://ui.adsabs.harvard.edu/abs/2025SusCS.11906076H}
}

@ARTICLE{DIsanto2016,
       author = {{D'Isanto}, A. and {Cavuoti}, S. and {Brescia}, M. and {Donalek}, C. and {Longo}, G. and {Riccio}, G. and {Djorgovski}, S.~G.},
        title = "{An analysis of feature relevance in the classification of astronomical transients with machine learning methods}",
      journal = {\mnras},
         year = 2016,
        month = apr,
       volume = {457},
       number = {3},
        pages = {3119-3132},
          doi = {10.1093/mnras/stw157},
archivePrefix = {arXiv},
       eprint = {1601.03931},
 primaryClass = {astro-ph.IM},
       adsurl = {https://ui.adsabs.harvard.edu/abs/2016MNRAS.457.3119D}
}

\end{document}